\documentclass[aps,prl,twocolumn,superscriptaddress]{revtex4-1}

\usepackage{graphics}

\begin{document}


\title{Inverse-kinematics proton scattering on $^{50}$Ca: determining
  effective charges using complementary probes}


\author{L. A. Riley}
\affiliation{Department of Physics and Astronomy, Ursinus College,
  Collegeville, PA 19426, USA}

\author{M. L. Agiorgousis} \affiliation{Department of Physics and
  Astronomy, Ursinus College, Collegeville, PA 19426, USA}

\author{T.R. Baugher} \affiliation{National Superconducting Cyclotron
  Laboratory, Michigan State University, East Lansing, MI, 48824, USA}

\author{D. Bazin} \affiliation{National Superconducting Cyclotron
  Laboratory, Michigan State University, East Lansing, MI, 48824, USA}

\author{M. Bowry} \affiliation{National Superconducting Cyclotron
  Laboratory, Michigan State University, East Lansing, MI, 48824, USA}

\author{P. D. Cottle} \affiliation{Department of Physics, Florida
  State University, Tallahassee, FL 32306, USA}

\author{F. G. DeVone} \affiliation{Department of Physics and
  Astronomy, Ursinus College, Collegeville, PA 19426, USA}

\author{A. Gade} \affiliation{National Superconducting Cyclotron
  Laboratory, Michigan State University, East Lansing, MI, 48824, USA}

\author{M. T. Glowacki} \affiliation{Department of Physics and
  Astronomy, Ursinus College, Collegeville, PA 19426, USA}

\author{K. W. Kemper} \affiliation{Department of Physics, Florida
  State University, Tallahassee, FL 32306, USA}

\author{E. Lunderberg} \affiliation{National Superconducting Cyclotron
  Laboratory, Michigan State University, East Lansing, MI, 48824, USA}

\author{D. M. McPherson} \affiliation{Department of Physics, Florida
  State University, Tallahassee, FL 32306, USA}

\author{S. Noji} \affiliation{National Superconducting Cyclotron
  Laboratory, Michigan State University, East Lansing, MI, 48824, USA}

\author{F. Recchia} \altaffiliation{Dipartimento di Fisica e
  Astronomia “Galileo Galilei”, Universit‘a degli Studi di Padova,
  I-35131 Padova, Italy} \affiliation{National Superconducting
  Cyclotron Laboratory, Michigan State University, East Lansing, MI,
  48824, USA}

\author{B. V. Sadler} \affiliation{Department of Physics and
  Astronomy, Ursinus College, Collegeville, PA 19426, USA}

\author{M. Scott} \affiliation{National Superconducting Cyclotron
  Laboratory, Michigan State University, East Lansing, MI, 48824, USA}

\author{D. Weisshaar} \affiliation{National Superconducting Cyclotron
  Laboratory, Michigan State University, East Lansing, MI, 48824, USA}

\author{R. G. T. Zegers} \affiliation{National Superconducting
  Cyclotron Laboratory, Michigan State University, East Lansing, MI,
  48824, USA} \affiliation{Joint Institute for Nuclear Astrophysics,
  Michigan State University, East Lansing, MI 48824, USA}


\date{\today}

\begin{abstract}
We have performed measurements of the $0_\mathrm{g.s.}^+ \rightarrow
2_1^+$ excitations in the neutron-rich isotopes $^{48,50}$Ca via
inelastic proton scattering on a liquid hydrogen target, using the
GRETINA $\gamma$-ray tracking array.  A comparison of the present
results with those from previous measurements of the lifetimes of
the $2_1^+$ states provides us the ratio of the neutron and proton
matrix elements for the $0_\mathrm{g.s.}^+ \rightarrow 2_1^+$
transitions.  These results allow the determination of the ratio of
the proton and neutron effective charges to be used in shell model
calculations of neutron-rich isotopes in the vicinity of $^{48}$Ca.
\end{abstract}


\maketitle

Isotopes within a few nucleons of the doubly-magic nuclei provide the
foundation for the nuclear shell model.  The simplicity of the
wavefunctions of the valence nucleons in these nuclei allows the
determination of the two-body matrix elements necessary for predicting
the energies of excited states and the effective charges used in
calculating transition strengths in a given shell-model space.
Effective charges reflect the strength of the coupling between the
motion of the valence nucleons and the virtual excitations of the core
nucleons.

The neutron-rich exotic calcium isotopes beyond the doubly-magic nucleus
$^{48}$Ca are particularly interesting because of shell effects caused
by the filling of the $p_{3/2}$ and $p_{1/2}$ neutron orbits that are
just above the major $N=28$ shell closure.  As a result, the
determination of the effective charges for these neutron-rich calcium
isotopes is of particular interest.  Valiente-Dob\'{o}n et
al. \cite{Val09} addressed this by determining the electromagnetic
strengths of the $2_1^+ \rightarrow 0^+_\mathrm{g.s.}$ transition in
$^{50}$Ca and the transition between the ground state and 1065 keV
$11/2^-$ state in $^{51}$Sc via lifetime measurements.  They examined
two candidate sets of effective charges, one of which was recommended
in an extensive shell model study of the $fp$ shell \cite{Hon04} and
another of which was extracted from an experimental study of the
mirror nuclei $^{51}$Fe and $^{51}$Mn \cite{duR04}.  On the basis of
their measurements, Valiente-Dob\'{o}n {\it et al.} argued for the set
from Ref. \cite{Hon04}.

However, measurements of electromagnetic strengths only determine the
proton transition matrix elements and do not determine the neutron
transition matrix elements, which are particularly important in the
case of $^{50}$Ca, because the only valence nucleons are neutrons when
assuming an inert $Z = 20$ proton core.  Here we report the results of
a measurement of the $0^+_\mathrm{g.s.}  \rightarrow 2_1^+$ excitation
in $^{50}$Ca using the inelastic scattering of protons, for which the
transition matrix element includes contributions from both protons and
neutrons.  The experiment was performed in inverse kinematics with a
beam of radioactive $^{50}$Ca ions.  By comparing the results of the
present proton-scattering measurement with the previous measurement of
the electromagnetic matrix element deduced from the lifetime
measurement of Ref.~\cite{Val09}, we can determine the neutron
transition matrix element and demonstrate that the proton effective
charge must be approximately a factor of 3 larger than the neutron
effective charge, which is consistent with the effective charges
proposed by Honma et al. \cite{Hon04}.  The determination of the
neutron transition matrix element via the present proton scattering
measurement significantly strengthens the assertion given in
Ref.~\cite{Val09} that the effective charges proposed by Honma et
al. are valid near $^{48}$Ca.  We also report a similar measurement of
$^{48}$Ca for which the proton-scattering deformation length has
already been measured.~\cite{Gru72, Fuj88, Set85}

The experiment was performed at the Coupled-Cyclotron Facility of the
National Superconducting Cyclotron Laboratory at Michigan State
University. A cocktail beam was produced by the fragmentation of a
130~MeV/u $^{76}$Ge primary beam in a 376~mg/cm$^2$ $^9$Be production
target. The secondary products were separated by the A1900 fragment
separator~\cite{A1900}. A 45~mg/cm$^2$ aluminum achromatic wedge was
used to enhance separation of the cocktail by $Z$.

Secondary beam particles were identified upstream of the reaction
target by energy loss in a silicon pin diode located at the object of
the analysis line of the S800 magnetic spectrograph~\cite{S800} and by
time of flight from the A1900 extended focal plane to the S800 object
scintillator.  The beam then traversed the Ursinus College Liquid
Hydrogen Target. The target, based on the design of Ryuto et
al.~\cite{Ryu05}, was installed at the pivot point of the
S800. Beam-like reaction products were identified by energy loss in
the S800 ion chamber and time of flight from the S800 object
scintillator to a scintillator in the focal plane of the S800.  The
secondary beam contained components spanning the range $14 \leq Z \leq
23$, including $^{48, 50}$Ca, the subjects of the present work.  A
total of $4.3 \times 10^8$ secondary beam particles traversed the
reaction target in 114 hours. Of these, $3.0 \times 10^7$ particles
were tagged as $^{50}$Ca particles in both incoming and outgoing
particle identification gates, and $4.0 \times 10^6$ particles were
tagged as $^{48}$Ca, corresponding to 73~particles per second and
9.3~particles per second, respectively.

The liquid hydrogen was contained in a 30~mm thick cylindrical
aluminum target cell with 125~$\mu$m Kapton entrance and exit
windows. The cell was surrounded by a 1~mm thick aluminum radiation
shield with entrance and exit windows covered by 5~$\mu$m aluminized
Mylar foil. The temperature and pressure of the target cell were
maintained at 16.0~K and 868~Torr throughout the experiment. The
Kapton windows were deformed by the pressure differential between the
cell and the vacuum, adding to the nominal 30~mm target thickness.

The GRETINA~\cite{GRETINA} $\gamma$-ray tracking array, consisting of
28 36-fold segmented high purity germanium crystals packaged in seven
clusters of four crystals each, was installed at the pivot point of
the S800 in a configuration designed to accommodate the liquid
hydrogen target. Two of the seven modules were mounted at 58$^\circ$,
four at 90$^\circ$, and one at 122$^\circ$ with respect to the beam
axis.

We used realistic \textsc{geant4}~\cite{GEANT4} simulations to extract
$\gamma$-ray yields from measured spectra.  The simulation code
includes detailed models of GRETINA and the liquid hydrogen target to
account for scattering of $\gamma$ rays by the dead material
surrounding the liquid hydrogen and the active detector volumes.  The
code reproduces measured photopeak efficiencies within 5\% between
300~keV and 3.5~MeV.  In addition to $\gamma$ rays, simulated beam
particles are also tracked as they traverse the target and undergo
$\gamma$ decay in flight.  The Kapton entrance and exit windows of the
target cell bulged outward due to the 868~Torr pressure differential
between the liquid hydrogen and the evacuated beamline. Therefore, the
roughly 4~mm $\times$ 8~mm beam spot sampled a range of target
thicknesses.  We varied the thickness of the simulated window bulges
to fit the kinetic energy distributions of the outgoing scattered beam
particles measured with the S800 to determine an effective target
thickness of 34.3~mm, corresponding to an areal density of
258~mg/cm$^2$.  The \textsc{geant4} simulations described above were
used to determine that inelastic scattering reactions took place at an
average $^{48}$Ca beam energy of 98~MeV/u and an average $^{50}$Ca
energy of 90~MeV/u. The corresponding beam velocities, $v/c = 0.43$
and $v/c = 0.41$, were used in the Doppler reconstruction of the
measured and simulated $\gamma$-ray spectra.

\begin{figure}
\scalebox{0.585}{
  \includegraphics{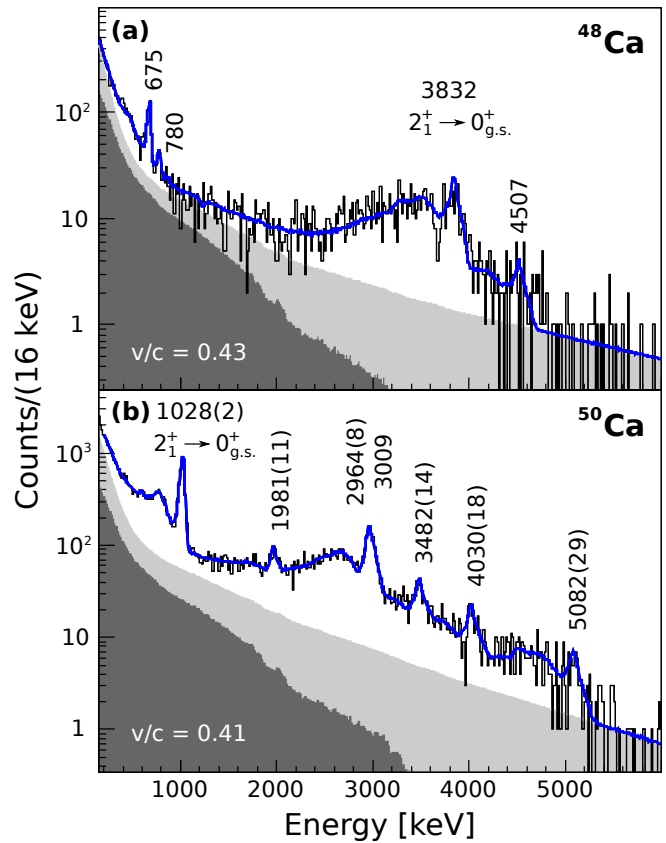}
}
\caption{\label{50ca_spectrum} (Color online) Doppler-corrected
  spectra of gamma rays measured in coincidence with incoming and
  outgoing (a) $^{48}$Ca and (b) $^{50}$Ca particles. The solid curves
  are the \textsc{geant4} fits described in the text. The shaded
  region is the background, consisting of non-prompt (dark gray) and
  prompt (light gray) components.}
\end{figure}

The Doppler-corrected $\gamma$-ray spectra collected in coincidence
with inelastically scattered $^{48}$Ca and $^{50}$Ca beam particles
are shown in Figure~\ref{50ca_spectrum}.  The solid curve in each
panel is a fit of a linear combination of \textsc{geant4} simulations
of the response of GRETINA to the observed $\gamma$ rays and prompt
and non-prompt background components, shaded in light gray and dark
gray in the figure, respectively. The non-prompt
background is due to random coincidences between scattered beam
particles and the room background. We fit this component of the
spectrum with a scaled room background measurement. The scaling is
determined with a fit to the laboratory-frame spectrum and fixed in
the projectile-frame fit.  The remaining prompt component of the
background, shaded in light gray in Figure~\ref{50ca_spectrum}, can be
successfully accounted for by including two exponential functions in
the projectile-frame fit. The $\gamma$-ray energies and intensities
extracted from the fits are listed in Table~\ref{tab:gammas} along
with branching ratios.

In the case of $^{48}$Ca, we included in the fit all of the gamma rays
de-exciting states up to an excitation energy of 6~MeV that are
known~\cite{Gru72, Fuj88, Set85} to be strongly-populated by proton
scattering.  We did not have sufficient statistics to independently
determine the energies of the 648~keV, 1538~keV, and 4507~keV gamma
rays, and the 648/675~keV, 754/758~keV, and 863/867~keV doublets in
the $^{48}$Ca spectrum were not well resolved in the measured
spectrum. With the exception of the relatively intense 675~keV
transition, we fixed the energies of these transitions at the adopted
values~\cite{Bur06} in the fit. Similarly, in the $^{50}$Ca spectrum,
the 2964/3009 keV doublet is not resolved. We used the adopted
energy~\cite{Ele11} for the much fainter 3009~keV member of the
doublet in the fit.

We observe a $\gamma$-ray at 5082(29)~keV in $^{50}$Ca that has not been
previously observed. We do not have sufficient statistics to place it
in the level scheme on the basis of $\gamma$-$\gamma$ coincidence
gating. The neutron separation energy in $^{50}$Ca is
6353(8)~keV~\cite{Ele11}.  We therefore suspect that the 5082~keV
$\gamma$ ray feeds either the $2^+_1$ state at 1027~keV or the ground
state. Although its energy is compatible with a $\gamma$ ray
de-exciting the known state at 5110~keV, the tentative assignment of
$J^\pi = 5^-$ to this state suggests that it is unlikely to decay
directly to the ground state.

\begin{table*}
\caption{\label{tab:gammas} Level energies, spins and parities, and
  $\gamma$-ray energies from Refs.~\cite{Bur06, Ele11}, $\gamma$-ray
  energies, intensities relative to that of the $2^+_1 \rightarrow
  0^+_\mathrm{g.s.}$ transitions, branching ratios (BR), and cross
  sections from the present work.}
\begin{ruledtabular}
\begin{tabular}{crccccccc}
&\multicolumn{3}{c}{Refs.~\cite{Bur06,Ele11}} 
                       & & \multicolumn{4}{c}{Present work}\\
\cline{2-5}\cline{6-9}
&$E_\mathrm{level}$ [keV] & $J^\pi$ [$\hbar$] & $E_\gamma$ [keV] &BR [\%]&
  $E_\gamma$ [keV] & $I_\gamma$ [\%] & BR [\%] & $\sigma$ [mb] \\\hline\hline
$^{48}$Ca&
3831.72(6)&$2^+$         &3832.2(5)&        &3842(12)&100(5) &  & 5.2(19)\\
&4506.78(5)&$3^-$        &675.05(7)&80.0(14)& 678(2)& 45(6) & 72(7) & 6.8(14)\\
&          &             &4507.3(5)&20.0(16)&  4507 & 17(3) & 28(7) & \\
&4611.81(7)&$3(+)$       &780.25(15)&       & 782(8)& 12(3) &       & $<3.2$ \\
&5260.41(8)&$4(-)$       & 648.4(1)&   86(5)&   648 & 12(4) & 69(15)& $<1.9$ \\
&          &             & 753.8(1)&   14(5)&   754 &  5(3) & 31(15)& \\
&5369.59(6)&$3-$         & 757.5(1)&        &   758 &   --    &     & $<1.1$ \\
&          &             & 862.7(1)&        &   863 & 4.7(30) &     & \\
&          &             & 866.9(1)&        &   867 & $<3$    &     & \\
&          &             &1537.8(1)&        &  1538 & 2.4(30) &     & \\\hline
$^{50}$Ca&
1026.7(1)&$2^+$          &1026.7(1)&& 1028(2) &100(5)  &  & 3.4(11)\\
&3002.1(5)&($2^+$)       &1975.3(5)&& 1981(11) & 9(1)  &  & 1.4(2)\\
&3997.1(2)&($3^-$)       &2970.2(2)&& 2964(8)  &42(4)  &  & 6.8(6)\\
&4035.7(4)&($1^+$, $2^+$)&4035.6(5)&  62(1) & 4030(18) & 12(3)& 44(12) & 3.5(7)\\
&         &              &3008.9(5)&  38(2) &    3009  & 9(3)  & 56(15) &\\
&4515.0(1)&($4^+$)       &3488.2(1)&& 3482(14) &13(1)  &  & 1.6(2)\\
&5109.8(2)&($5^-$)       & 594.8(1)&&  603(11) & 2.6(9)& & 0.4(1)\\\cline{2-9}
&         &              &         && 5082(29) &5.4(6) & & \\
\end{tabular}
\end{ruledtabular}
\end{table*}

A 22-hour empty-cell measurement was also made to study possible
contamination of the $\gamma$-ray spectrum by reactions in the Kapton
windows and aluminized Mylar foils. No discernible peaks were observed
in the empty-cell spectra collected in coincidence with scattered
$^{48}$Ca and $^{50}$Ca particles. 

The uncertainties in both the position and velocity of the scattered
beam particles undergoing $\gamma$ decay in the thick target lead to
significant Doppler broadening of the photopeaks in the
Doppler-corrected spectrum. The photopeak of a $\gamma$ ray
de-exciting a state with a lifetime on the order of hundreds of
picoseconds or more is further broadened due to the extended range of
decay positions. The photopeak of the 1027~keV $\gamma$ ray in the
spectrum of $^{50}$Ca is slightly broadened relative to a simulated
peak assuming zero lifetime for the $2^+_1$ state. A best fit to the
spectrum is obtained with a mean lifetime of $\tau = 99(8)$~ps, in
good agreement with the recoil distance Doppler shift method result
from Ref.~\cite{Val09} of $\tau = 96(3)$~ps for the $2^+_1$ state of
$^{50}$Ca.

\begin{figure*}
\scalebox{0.4}{
  \includegraphics{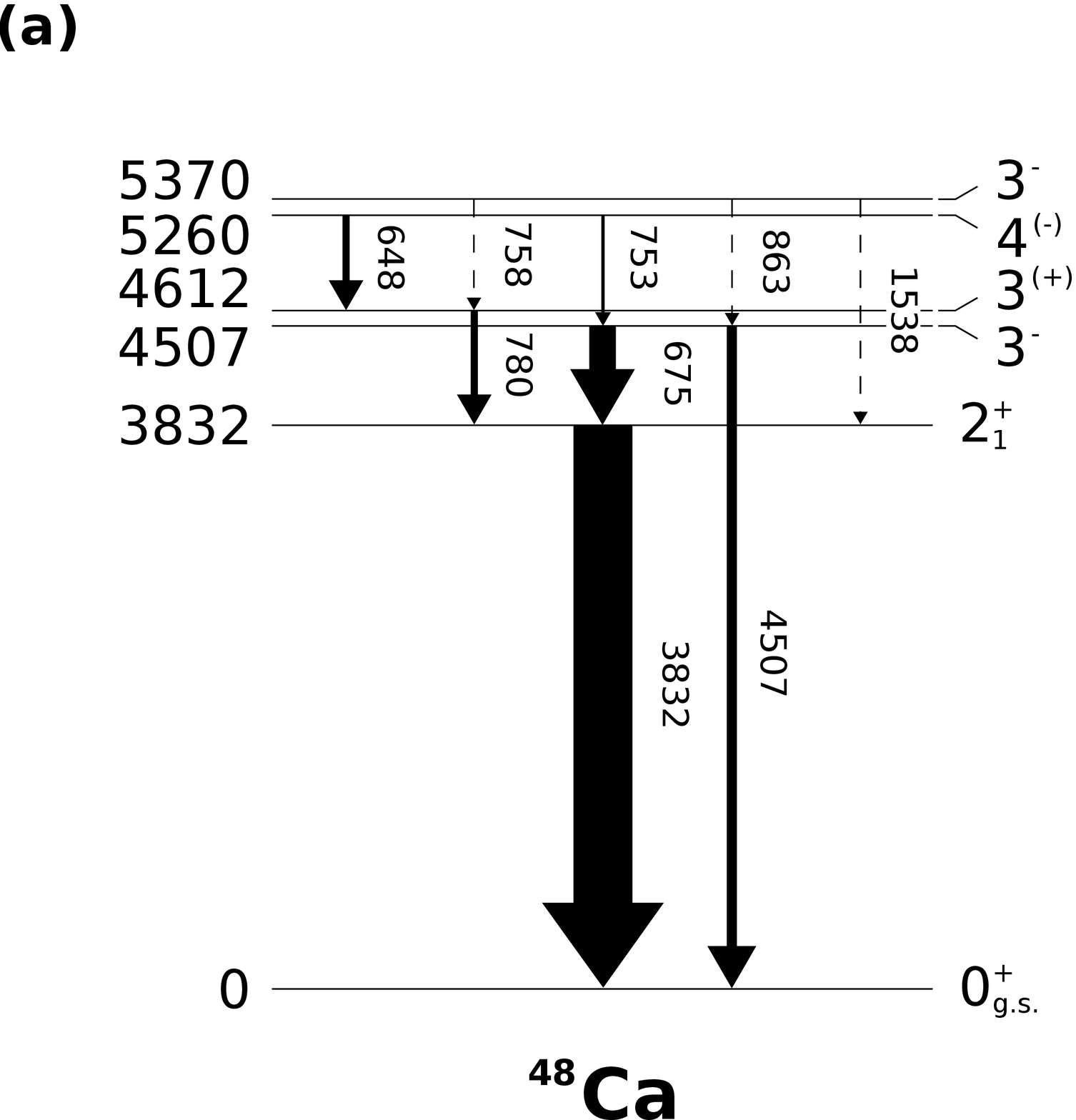}
  \hspace*{1.5 cm}
  \includegraphics{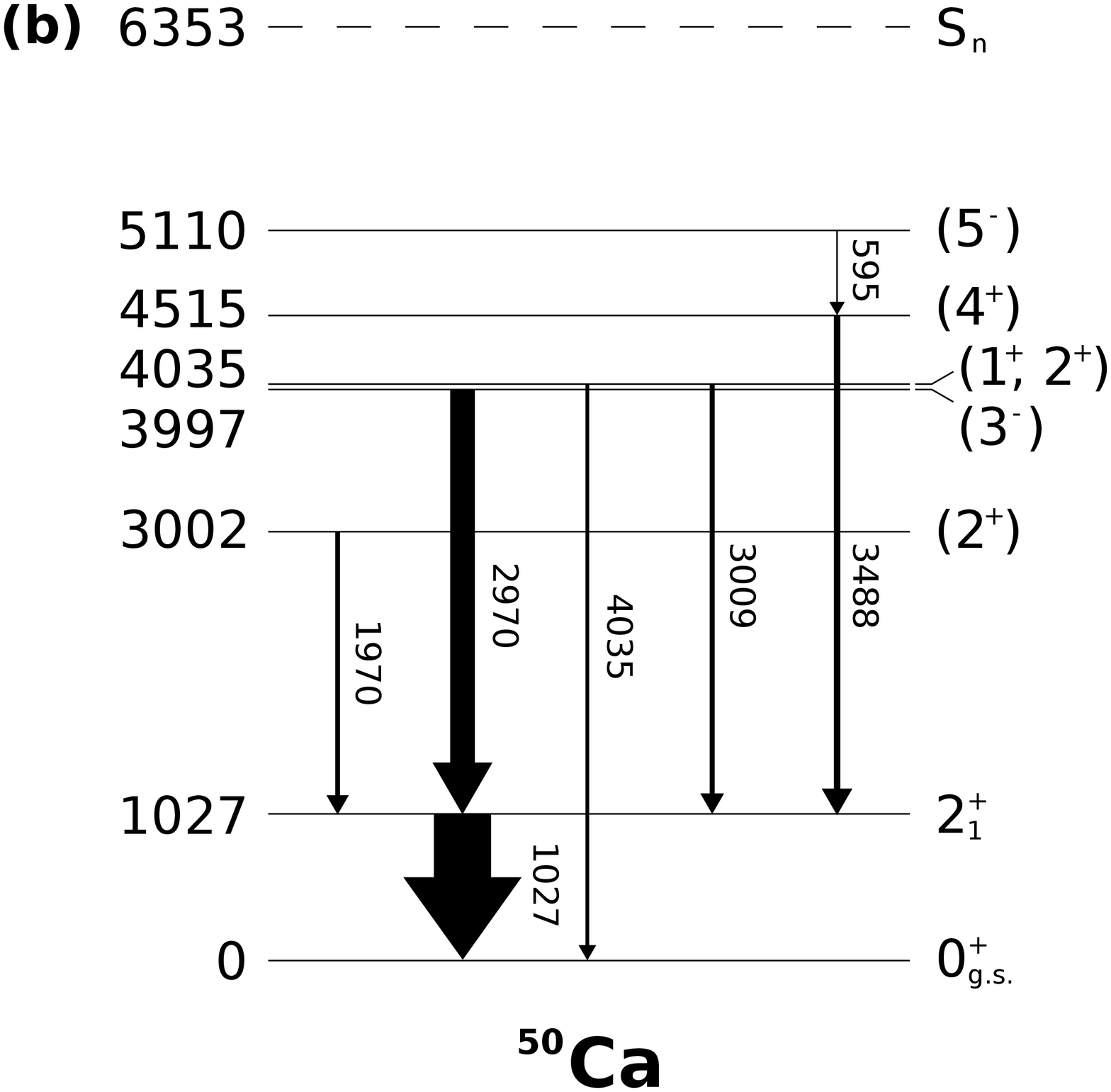}
}
\caption{\label{ca50_level}  Partial level schemes of (a) $^{48}$Ca
  from Ref.~\cite{Bur06} and (b) $^{50}$Ca from Ref.~\cite{Ele11},
  showing levels populated in the present work. Arrow widths are
  proportional to the measured $\gamma$-ray intensities.} 
\end{figure*}

Partial level schemes of $^{48}$Ca and $^{50}$Ca are shown in
Figure~\ref{ca50_level}.  The established level schemes enable us to
apply feeding corrections to the $\gamma$-ray cross sections to deduce
the cross sections given in Table~\ref{tab:gammas} for direct
population of excited states via proton scattering. In the error range
of the $2^+_1$-state cross section in $^{50}$Ca, we have accounted for
the possibility that the unplaced 5082~keV $\gamma$ ray feeds the
$2^+_1$ state.

We used the coupled-channels code \textsc{ECIS95}~\cite{Ray95} to
extract proton-scattering deformation lengths from our measured
cross sections for inelastic scattering to the $2^+_1$ states of
$^{48,50}$Ca. Using the global optical potential of Ref.~\cite{Kon03}
and treating the $2^+_1$ states as quadrupole vibrations
yields proton-scattering deformation lengths $\delta_{2 \, (p,p')} =
0.78(11)$~fm for $^{48}$Ca and $0.57(9)$~fm for $^{50}$Ca. 
Using a rotational model for $^{50}$Ca yields $\delta_{2 \, (p,p')} =
0.56(9)$~fm. The discrepancy between the vibrational- and
rotational-model results is negligible compared with the experimental
uncertainty.

The deformation length of the $2^+_1$ state of $^{48}$Ca has been
determined via distorted-wave Born approximation (DWBA) analyses of
proton angular distributions over a broad range of proton energies ---
$\delta_{2 \, (p,p')} = 0.70(3)$~fm at 40 MeV~\cite{Gru72},
$0.61(3)$~fm at 65 MeV~\cite{Fuj88}, and $0.61(2)$~fm at
500~MeV.~\cite{Set85}

The ratio of neutron to proton transition matrix elements, $M_n$ and
$M_p$, is related to the corresponding proton and neutron deformation
lengths, $\delta_n$ and $\delta_p$, by~\cite{Ber83} 
\begin{equation}
\label{eq:MnMp1}
\frac{M_n}{M_p} = \frac{N \delta_n}{Z \delta_p}.
\end{equation}
The lifetimes of the $2^+_1$ states of $^{48,50}$Ca are
known~\cite{Bur06, Val09} to be and $\tau = 38.7(19)$~fs and
$96(3)$~ps. These purely electromagnetic measurements give the proton
deformation lengths $\delta_p = 0.479(13)$~fm and $0.304(4)$~fm
directly.  Proton scattering is sensitive to protons as well as to
neutrons. Hence the proton-scattering deformation lengths from the
present work can be combined with the proton deformation lengths to
determine $M_n/M_p$ for the $2^+_1$ states.  The deformation length
measured using a probe $F$ with a mixed sensitivity to protons and
neutrons is related to the neutron and proton deformation lengths
via~\cite{Ber81}
\begin{equation}
\label{eq:dF}
\delta_F = \frac{b_n^F N \delta_n + b_p^F Z \delta_p}
                          {b_n^F N + b_p^F Z},
\end{equation}
where $b_n^F$ ($b_p^F$) are the external-field neutron (proton)
interaction strengths of the probe. In the case of a pure
electromagnetic probe combined with proton scattering,
Eqs.~(\ref{eq:MnMp1}) and (\ref{eq:dF}) yield 
\begin{equation}
\label{eq:MnMp2}
\frac{M_n}{M_p} = \frac{b_p}{b_n} 
\left( \frac{\delta_{(p,p')}}{\delta_p}
\left( 1 + \frac{b_n}{b_p} \frac{N}{Z} \right) - 1 \right).
\end{equation}
The ratio $b_n/b_p$ is approximately 3 for proton scattering below
50~MeV and approximately 1 at 1~GeV~\cite{Ber81}, but it is not
well-determined at the intermediate beam energies of the present work. 

\begin{figure}
\scalebox{0.5}{
  \includegraphics{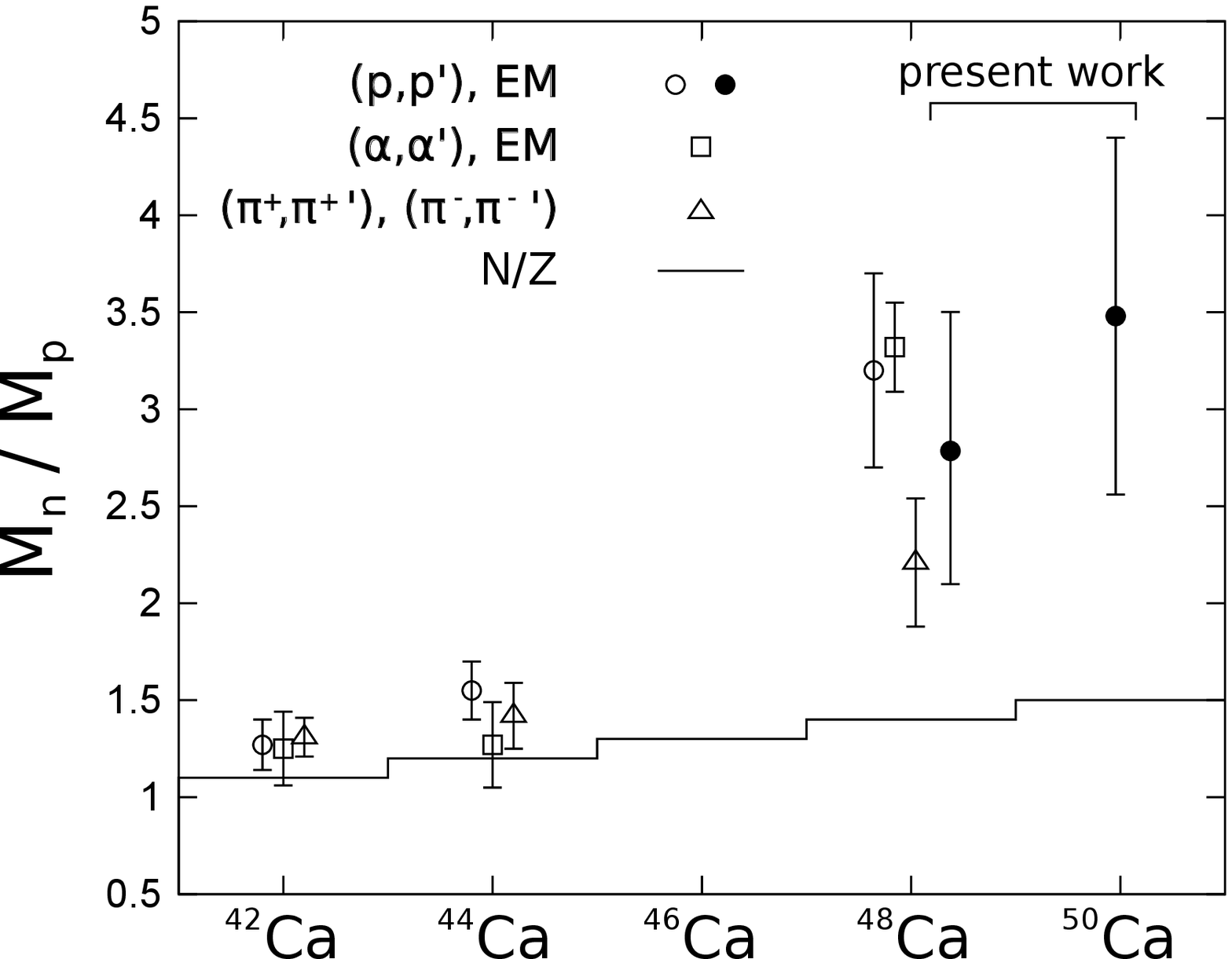}
}
\caption{\label{MnMp} The ratio of neutron to proton transition matrix
elements $M_n/M_p$ in the neutron-rich calcium isotopes. The open
symbols are taken from Ref.~\cite{Ber83} and the filled circles are
from the present work.}
\end{figure}

Values of $M_n/M_p$ for the $2^+_1$ state of $^{48}$Ca,
calculated using $b_n/b_p = 1$ and $b_n/b_p = 3$, are 2.5(4) and
2.9(6), respectively. These results span the range $2.1 \leq M_n/M_p
\leq 3.5$. Lacking a firm value of $b_n/b_p$, we adopt the result
$M_n/M_p = 2.8(7)$ to cover this range. For $^{50}$Ca, an identical
analysis gives $M_n/M_p = 3.5(9)$. These results along with the
$M_n/M_p$ values of $2^+_1$ states in the neutron-rich calcium
isotopes from Ref.~\cite{Ber83} are shown in Figure~\ref{MnMp}. 

In a nucleus like $^{50}$Ca that has only valence neutrons, the ratio
$M_n/M_p$ can be shown in a straightforward way to be equal to the
ratio of the shell model effective charges $e_p/e_n$ under a
reasonable assumption --- that the coupling between valence neutrons
and core  neutrons has the same strength as the coupling between
valence and core protons. We start with the formalism given by Brown
and Wildenthal \cite{Bro80} in which
\begin{equation}
\label{eq:MnMp3}
M_p = A_p (1 + \delta_{pp}) + A_n \delta_{pn},
\end{equation}
and
\begin{equation}
\label{eq:MnMp4}
M_n = A_n (1 + \delta_{nn}) + A_p \delta_{np}.
\end{equation}
Here, $A_p$ and $A_n$ are the proton and neutron strength amplitudes,
respectively, and the $\delta_{ab}$ account for the coupling of the
nucleons $b$ of the model space to the virtual excitations of the core
nucleons $a$.  In $^{50}$Ca, there are no valence protons so that $A_p
= 0$ and
\begin{equation}
\label{eq:MnMp5}
\frac{M_n}{M_p} = \frac{1 + \delta_{nn}}{\delta_{pn}}.
\end{equation}
If we assume that $\delta_{nn} = \delta_{pp}$ then the conventional model 
space proton effective charge $e_p$ is given by
\begin{equation}
\label{eq:MnMp6}
e_p = 1 + \delta_{pp} = 1 + \delta_{nn}.
\end{equation}
With the conventional definition of the neutron effective 
charge $e_n = \delta_{pn}$, we have
\begin{equation}
\label{eq:MnMp7}
\frac{e_p}{e_n} = \frac{M_n}{M_p}.
\end{equation}

Honma {\it et al.} \cite{Hon04} adopted $e_p=1.5$ and $e_n=0.5$, which
gives $e_p/e_n = 3.0$. In contrast, du Rietz {\it et al.} \cite{duR04}
concluded from their study of $^{51}$Fe and $^{51}$Mn that $e_p=1.15$
and $e_n=0.80$, giving $e_p/e_n = 1.4$. The present result, $e_p/e_n =
3.5(9)$, clearly favors the effective charges selected by Honma {\it
  et al.}, in agreement with Valiente-Dob\'{o}n {\it et al.}
\cite{Val09}.  This does not mean that the conclusion of du Rietz {\it
  et al.} regarding effective charges is incorrect.  Instead, it
indicates that the effective charges in the neutron-rich isotopes in
the vicinity of $^{48}$Ca are different than they are near the $N
\approx Z$ nuclei, $^{51}$Fe and $^{51}$Mn. Indeed, an isospin
dependence in effective charges was argued a long time ago by Bohr and
Mottelson in Ref.~\cite{Boh75}.

In summary, we have performed a measurement of inelastic proton
scattering in inverse kinematics on the neutron-rich isotopes
$^{48,50}$Ca.  Through a comparison of the proton-scattering
deformation lengths from the present work with previous lifetime
measurements, we determined the ratio of the neutron and proton matrix
elements for the $0_\mathrm{g.s.}^+ \rightarrow 2_1^+$ transitions in
these nuclei.  The $M_n/M_p$ value for $^{48}$Ca is in good agreement
with prior results. Our result for $^{50}$Ca allows us for the first
time to determine the ratio of the proton and neutron effective
charges in the $fp$ shell for isotopes in the vicinity of $^{48}$Ca
and to conclude that the effective charges adopted for these nuclei by
Honma {\it et al.} \cite{Hon04} are appropriate for use in shell model
calculations in this region.

\begin{acknowledgments}
We are grateful for conversations with B.A. Brown. We also thank
T.J. Carroll for the use of the Ursinus College Parallel Computing
Cluster, which is supported by NSF grant no. PHY-1205895. This work
was supported by the National Science Foundation under Grant Nos.
PHY-1303480, PHY-1064819, and PHY-1102511. GRETINA was funded by the
US DOE - Office  of Science. Operation of the array at NSCL is
supported by NSF under Cooperative Agreement PHY-1102511(NSCL) and DOE
under grant DE-AC02-05CH11231(LBNL).
\end{acknowledgments}

\end{document}